\documentclass[onecolumn,aps,showpacs,superscriptaddress]{elsart}
\usepackage{graphicx}
\usepackage{dcolumn}
\usepackage{bm}
\begin{document}

\def\sNN{$\sqrt{s_{_{NN}}}$}
\def\pT{$p_T$ }
\def\geqslant{${}^{>}_{_}$}
\def\geqsim{${}^{>}_{\sim}$}
\def\Journal#1#2#3#4{{#1} {\bf #2} (#4) #3}
\def\NCA{\em Nuovo Cimento}
\def\NIM{\em Nucl. Instr. Meth.}
\def\NIMA{{\em Nucl. Instr. Meth.} A}
\def\NPB{{\em Nucl. Phys.} B}
\def\NPA{{\em Nucl. Phys.} A}
\def\PLB{{\em Phys. Lett.}  B}
\def\PRL{{\em Phys. Rev. Lett.}}
\def\PRC{{\em Phys. Rev.} C}
\def\PRD{{\em Phys. Rev.} D}
\def\ZPC{{\em Z. Phys.} C}
\def\JPG{{\em J. Phys.} G}
\def\EPJ{{\em Eur. Phys. J.} C}
\def\RPP{{\em Rep. Prog. Phys.}}

\begin{frontmatter}


\title{Extensive Particle Identification with TPC and TOF at the STAR Experiment}


\author[USTC,BNL]{Ming Shao},
\author[PURDUE]{Olga Barannikova},
\author[USTC,LBNL]{Xin Dong},
\author[BNL]{Yuri Fisyak},
\author[USTC,BNL]{Lijuan Ruan},
\author[LBNL]{Paul Sorensen},
\author[BNL]{Zhangbu Xu}


\address[USTC]{University of Science and Technology of China, Hefei, Anhui 230026, China}
\address[BNL]{Brookhaven National Laboratory, Upton, New York 11973, USA}
\address[LBNL]{Lawrence Berkeley National Laboratory, Berkeley, California 94720, USA}
\address[PURDUE]{Physics Department, Purdue University, West
Lafayette, Indiana 47907, USA}

\begin{abstract}
Particle identification (PID) capabilities are studied by using
the Time Projection Chamber (TPC) and a Time-Of-Flight (TOF)
detector together at STAR. The identification capability of
charged hadrons is greatly extended compared with that achieved by
TPC and TOF separately. Particle spectra from p+p, d+Au collisions
at $\sqrt{s_{_{NN}}}=$200 GeV and Au+Au collisions at
$\sqrt{s_{_{NN}}}=$62.4 GeV are used to develop the methods. The
transverse momentum ($p_T$) ranges of $\pi$, and $p(\bar{p})$
identification are from $\sim0.3$ GeV/$c$ to $\sim10$ GeV/$c$. The
high $p_T$ reach is limited by statistics in current data sets. An
important conceptual advance was developed to identify electrons
by using a combination of dE/dx in TPC and velocity information
from the TOF detectors, which is important for future low-mass
dilepton program at STAR.
\end{abstract}

\begin{keyword}
particle identification, TOF, TPC, dE/dx, STAR
\PACS 29.40.Cs, 29.40.Gx, 29.85.+c
\end{keyword}
\end{frontmatter}
\maketitle

\section{Introduction}
One of the goals of the relativistic heavy ion program at RHIC is
to study quantum chromodynamics (QCD) at extreme condition
\cite{STARwhitepaper}. A unique strength of the solenoidal tracker
at RHIC (STAR)\cite{STARoverview} is its large, uniform acceptance
capable of measuring and identifying a substantial fraction of the
particles produced in heavy ion collisions. Detectors relevant to
the study presented in this article are the Time Projection
Chamber (TPC) \cite{TPC}, and a proposed barrel time-of-flight
(TOF) \cite{TOF}. For stable charged hadrons, the TPC provides
$\pi$/K ($\pi$+K/p) identification to $p_T \simeq$ 0.7 (1.1)
GeV/$c$ by the ionization energy loss (dE/dx) as usually been
quoted and presented in the physics analyses\cite{TPC}. However,
direct particle identification (PID) capability for stable hadrons
at intermediate/high $p_T$ is important for the study of
collective flow and strong early-stage interaction in the dense medium
formed in relativistic heavy ion collisions \cite{STARwhitepaper}.
STAR PID capability can be further enhanced by the proposed TOF. A
TOF system with a time resolution of ${}^{<}_{\sim}$ 100 ps at
STAR is able to identify $\pi$/K ($\pi$+K/p) to $p_T \simeq$ 1.6
(3.0) GeV/$c$, as demonstrated in Fig.~\ref{TOFrPID} (see also
\cite{TOFcronin}\cite{TOFHQ2004}).  In addition, with relativistic
rise of dE/dx from charged hadrons traversing the TPC at
intermediate/high $p_T$ (${}^{>}_{\sim}3$ GeV/$c$) and diminished
yields of electrons and kaons at this $p_T$ range, we can identify
pions and protons up to very high $p_T$ ($\simeq10$ GeV/$c$) in
p+p, p+A and A+A collisions at RHIC.  An important conceptual
advance was developed to identify electrons by using a combination
of dE/dx in the TPC and velocity information from the TOF detectors.
This has been used to measure charm yield via its semileptonic
decay \cite{TOFopencharm}. Electron identification and hadron
rejection power will be discussed in detail. This provides a
basic tool for the future dilepton measurements with the azimuthal
2$\pi$ coverage of a barrel time-of-flight system. The proposed
dilepton measurements will provide a penetrating probe into the
new state of dense matter produced in central heavy ion collisions
at RHIC, since leptons do not participate in strong interactions
occurring during hadronization and freeze-out.

\section{Experiment Setup}
At STAR\cite{STARoverview}, the main tracking device is a TPC,
covering full azimuthal angle and $\pm1.5$ units of pseudo-rapidity. A
dE/dx resolution of $\sim 8\%$ can be achieved by requiring the tracks
of charged particles to have at least 20 out of a maximum of 45 hits
in the TPC. Detailed descriptions of the TPC and its electronics
system have been presented in \cite{TPC}\cite{TPCFEE}.  One tray of
prototype time-of-flight detector based on multi-gap resistive plate
chambers \cite{MRPC} (TOFr) was installed in STAR in 2003. It covers
1/60 in azimuth and $-1 \leq \eta \leq 0$ in pseudo-rapidity at the
outer radius of the TPC, 220 cm from the interaction point. Two
identical pseudo-vertex position detectors (pVPD) were installed to
record the start time for the TOFr, each 5.4 m away from the TPC
center along the beam line. Each pVPD covers $\sim 19\%$ of the total
solid angle in $4.4 \leq |\eta| \leq 4.9$. More information of the
TOFr and the pVPD can be found in \cite{TOFr}\cite{TOFp}.  The data
used in this study were collected from Au+Au collisions at
$\sqrt{s_{NN}}=62.4$ GeV in 2004, and p+p and d+Au collisions at
$\sqrt{s_{NN}}=200$ GeV in 2003 at RHIC. The time resolution of the
TOF system is $\sim 105ps$ (Au+Au) and $\sim 120ps$ (d+Au)
respectively, which include pVPD's contribution of $\sim 55ps$ (Au+Au)
and $\sim 85ps$ (d+Au).

\section{Stable Hadron Identification}
\subsection{PID at intermediate/high $p_T$ by TPC}

\begin{figure}[ht]
\begin{center}
{\includegraphics[width=0.7\textwidth] {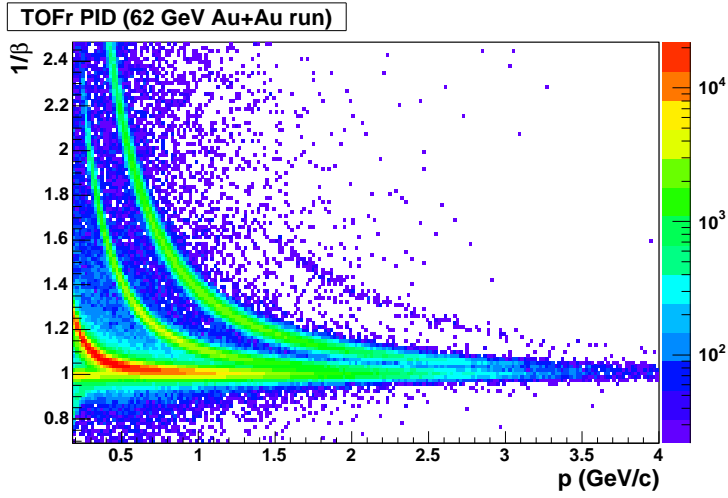}}
\end{center}
 \caption[]{$1/\beta$ vs. momentum for pions, kaons and (anti-)protons
from TOFr at 62.4 GeV Au+Au collisions. The separation between pions
and kaons ( (anti-)protons) is achieved to $p_T \sim$ 1.6 (3.0)
GeV/$c$..}
 \label{TOFrPID}
\end{figure}

\begin{figure}[ht]
\begin{center}
{\includegraphics[width=0.7\textwidth] {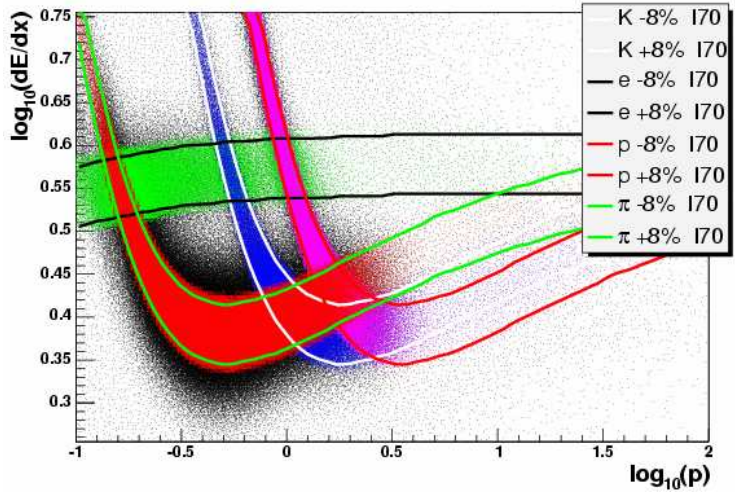}}
\end{center}
 \caption[]{Distribution of $log_{10}(dE/dx)$ as a function of
 $log_{10}(p)$ for electrons, pions, Kaons and (anti-)protons. The
 units of dE/dx and momentum ($p$) are keV/cm and GeV/$c$,
 respectively. The color bands denote within $\pm1\sigma$ the dE/dx
 resolution. {\it I70} means Bichsel's prediction for $30\%$
 truncated dE/dx mean. }
 \label{bigplot}
\end{figure}

\begin{figure}[ht]
\begin{center}
{\includegraphics[width=0.7\textwidth]
{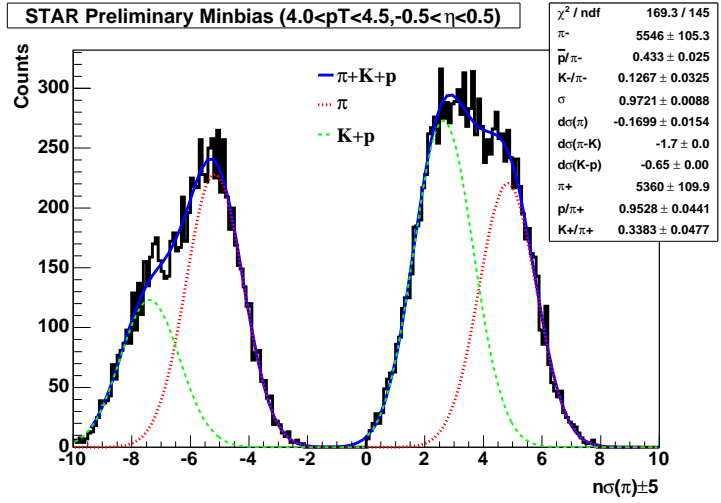}}
\end{center}
 \caption[]{dE/dx distribution normalized by pion dE/dx at $4<p_T<4.5$
GeV/$c$ and $|\eta| < 0.5$, and shifted by $\pm5$ for positive and
negative charged particles, respectively. The distribution is from
minimum-bias Au+Au collisions at $\sqrt{s_{_{NN}}}=62.4$ GeV.}
 \label{rdEdxpid}
\end{figure}

\begin{figure}[ht]
\begin{center}
{\includegraphics[width=0.7\textwidth]
{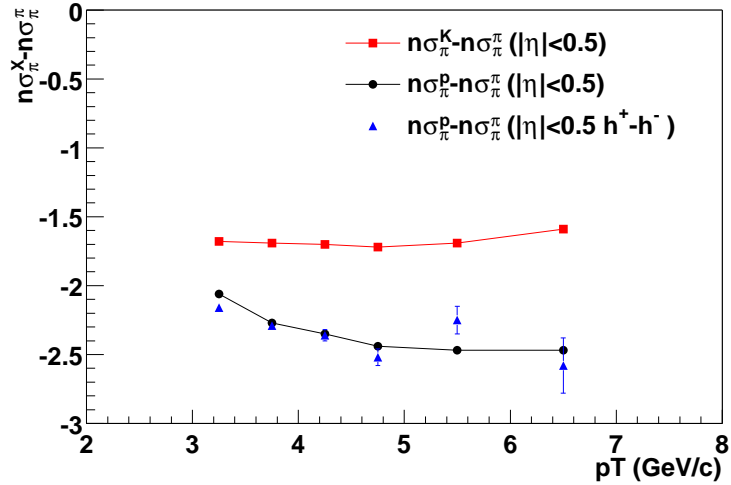}}
\end{center}
 \caption[]{The relative dE/dx peak position of K-$\pi$ (squares) and
p-$\pi$ (circles) in unit of standard resolution width
($\sigma_{\pi}$) of pion dE/dx as function of $p_T$. The triangles
are the peak positions of the dE/dx distribution of
$h^+-h^{-}=(p-\bar{p})+(K^+-K^-)+(\pi^+-\pi^-) \simeq (p-\bar{p})$.}
 \label{nsigchk}
\end{figure}
At $3 < p_T {}^{<}_{\sim} 10$ GeV/$c$, there is a difference of
about $15\%$ in the dE/dx between pions and kaons due to the pion
relativistic rise of the ionization energy loss. The difference
between that of pions and (anti-)protons is even larger. This
allows us to identify pions from other hadrons at this $p_T$ range
by the TPC alone at $2\sigma$ level.  The dE/dx resolution is
$\sim8\%$, as demonstrated in Fig.~\ref{bigplot}. Shown in
Fig.~\ref{rdEdxpid} is the $n\sigma_{\pi}$ distribution for
charged hadrons at $4 \leq p_T \leq 4.5$ GeV/$c$ and $|\eta| <
0.5$, where $n\sigma_{\pi}$ is the normalized dE/dx of pions. The
normalized dE/dx is defined by
$n\sigma_{X}^{Y}=log((dE/dx)_Y/B_{X})/\sigma_{X}$, in which $X,Y$
can be $e^{\pm},\pi^{\pm},K^{\pm}$ or $p(\bar{p})$. $B_{X}$ is the
expected mean dE/dx of a particle $X$, and $\sigma_{X}$ is the
dE/dx resolution of TPC. The $n\sigma_{\pi}^{\pi}$ distribution is
a normal Gaussian distribution with an ideal calibration. The
$n\sigma_{\pi}$ of positive and negative charged hadrons are
displaced by +5 and -5 respectively in Fig.~\ref{rdEdxpid}.
Fig.~\ref{nsigchk} shows the $p_T$ dependence of
$n\sigma_{\pi}^{K}$ and $n\sigma_{\pi}^{p(\bar{p})}$ relative to
$n\sigma_{\pi}^{\pi}$, as predicted by the Bichsel function for
the energy loss in thin layers of P10 \cite{TPC}.  From
Fig.~\ref{rdEdxpid} and measurement from TOF \cite{TOFHQ2004}, the
yield difference between positive and negative inclusive charged
hadrons is approximately that of proton and anti-proton (
$h^{+}-h^{-}=(p-\bar{p})+(K^{+}-K^{-})+(\pi^{+}-\pi^{-}) \simeq
(p-\bar{p})$, kaons' contribution is small). Therefore, the peak
positions of dE/dx distribution from $h^{+}-h^{-}$ should
represent well that of protons (since $n\sigma_{\pi}^{h^{+}-h^{-}}
\simeq n\sigma_{\pi}^{(p-\bar{p})} = n\sigma_{\pi}^{p(\bar{p})}$).
Indeed, the calibrated Bichsel function for protons matches the
dE/dx peak position of ($h^{+}-h^{-}$), as shown in
Fig.~\ref{nsigchk}. In addition, as shown later, the dE/dx
difference of protons and pions in the momentum range where PID
selection is possible by TOF is also found to be consistent with
the Bichsel function. These crosschecks confirm that the Bichsel
function can be used to constrain the relative dE/dx position
between kaons, (anti-)protons and pions. To extract pion yield, we
performed a six Gaussian fit to the dE/dx distributions of
positive and negative hadrons simultaneously as shown in
Fig.~\ref{rdEdxpid}.  The $n\sigma_{\pi}^{K}-n\sigma_{\pi}^{\pi}$
and $n\sigma_{\pi}^{p(\bar{p})}-n\sigma_{\pi}^{\pi}$ are fixed in
a six-Gaussian fit, where the six Gaussians are for $\pi^{\pm}$,
$K^{\pm}$ and $p(\bar{p})$ at a given $p_T$ bin. The sigma of the
six Gaussians are chosen to be the same. The pion yields extracted
from the fit can be found in \cite{rdEdx}. As shown by Fig.
~\ref{bigplot} and discussed in \cite{rdEdx}, the PID of pions at
$\sim3 < p_T < \sim$7-8 GeV/$c$ is obtained with this method. The
$p_T$ reach is limited by statistics and not by PID capability in
this study.
\begin{figure}[ht]
\begin{center}
{\includegraphics[width=0.7\textwidth]
{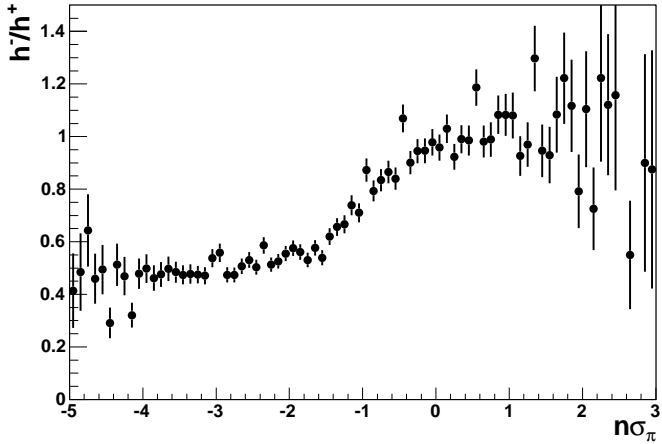}}
\end{center}
 \caption[]{$h^{-}/h^{+}$ vs dE/dx distribution normalized by pion
dE/dx at $3.5<p_T<4.0$ GeV/$c$ and $|\eta| < 0.5$. The distribution is
from minimum-bias Au+Au collisions at $\sqrt{s_{_{NN}}}=62.4$ GeV.}
 \label{hratio}
\end{figure}

 Fig.~\ref{nsigchk} shows that the dE/dx separation between kaon and
proton is less than one $\sigma$ at \pT between 3 and 5 GeV/$c$
and larger at \pT around 10 GeV/$c$. However, there are a few
methods we can use to identify protons and cross check the
contaminations from kaons. Fig.~\ref{hratio} shows the ratio of
negative hadrons over positive hadrons as a function of dE/dx in
unit of $n\sigma_{\pi}$ at $3.5<p_T<4.0$ GeV/$c$. Since the energy
loss of particles in TPC is independent of its charge sign, the
dependence of $h^{-}/h^{+}$ on $n\sigma_{\pi}$ is due to different
particle composition and due to the dE/dx separation between pion,
kaon and proton. Fig.~\ref{hratio} shows two plateaus. One from
$\bar{p}/p$ and the other from $\pi^{-}/\pi^{+}$. We will be able
to select protons with reasonable purity by requiring
$n\sigma_{\pi}<-2.5$ in this particular \pT bin. We can then use
an independent measurement of $K^{0}_{S}$ from V0 method to
constrain the charged kaon yields in the six-Gaussian fit or to
estimate the charged kaon contaminations described above to obtain
reliable proton yield. Detailed studies and measurements are
underway.

\subsection{Hadron PID at Intermediate $p_T$ by TPC+TOF}
\begin{figure}
\begin{minipage}{0.45\textwidth}
{\includegraphics[width=1.0\textwidth] {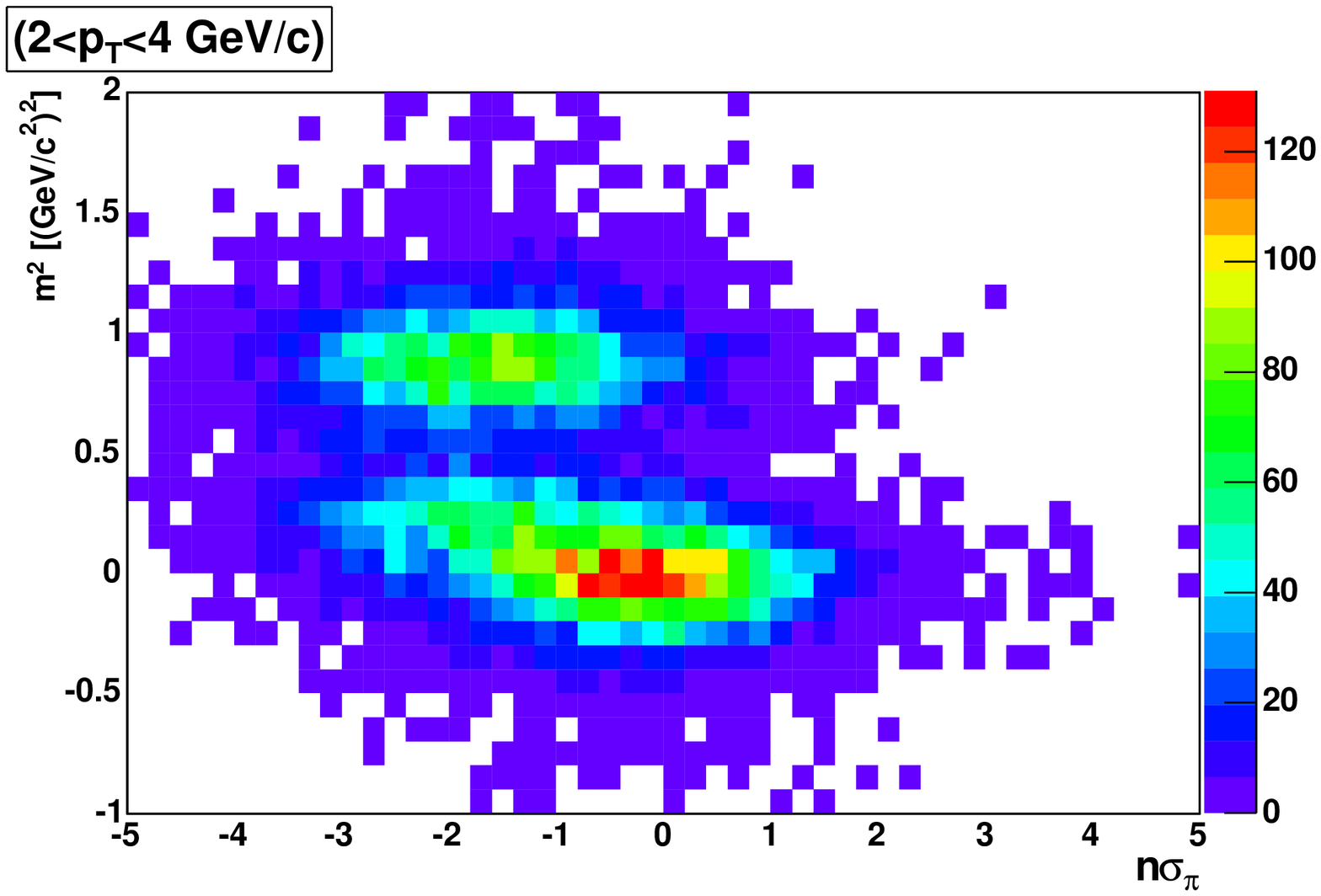}}
 \caption[]{Particle distribution as a function of $n\sigma_{\pi}$ and
 $m^2$ for $2<p_T<4$ GeV/$c$. The pion, kaon and (anti-)proton peaks
 can be clearly seen.}
 \label{nsigpi_m2}
\end{minipage}
\begin{minipage}{0.1\textwidth}
\end{minipage}
\begin{minipage}{0.45\textwidth}
{\includegraphics[width=1.0\textwidth] {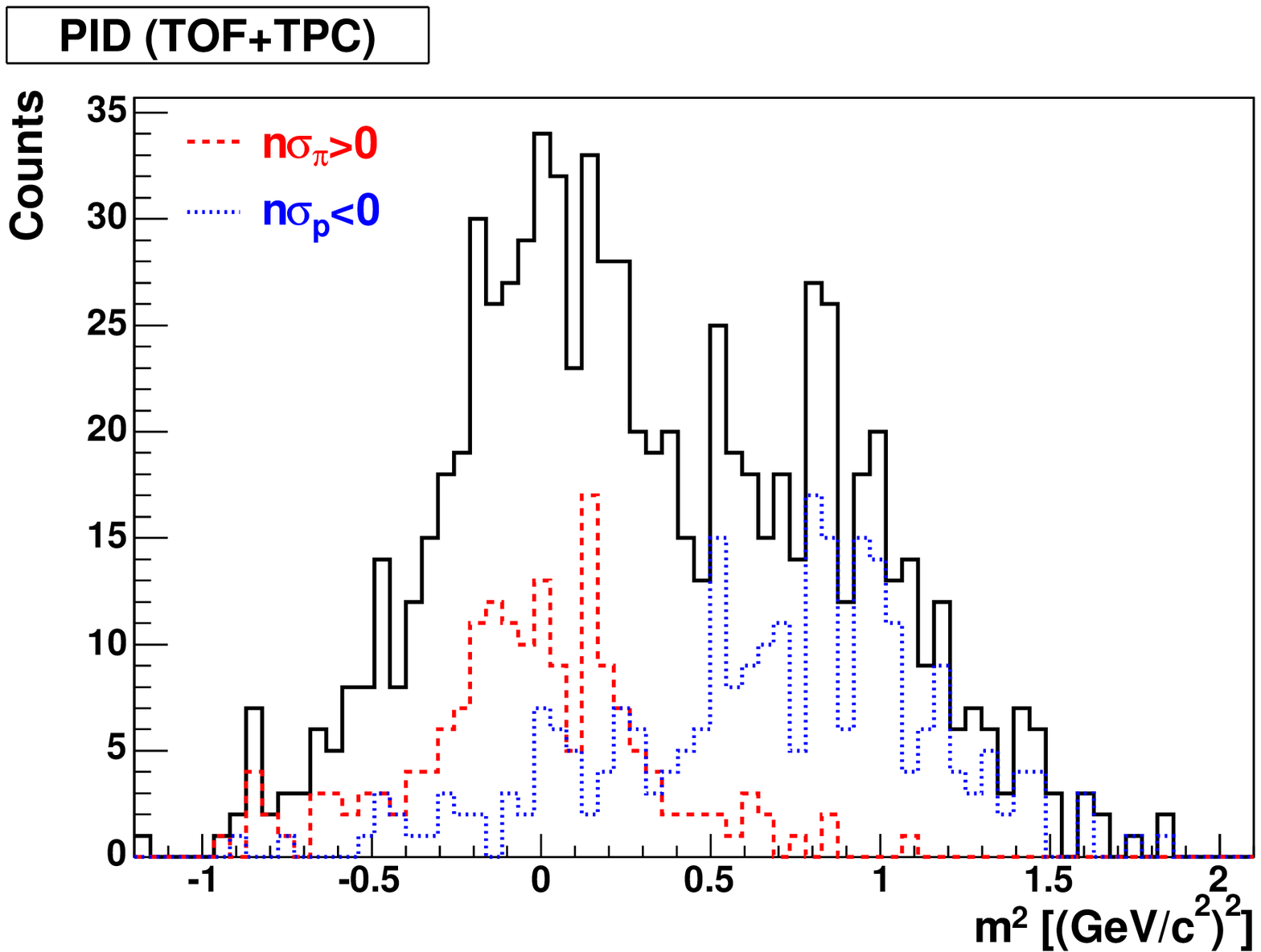}}
 \caption[]{$m^2$ distribution without dE/dx cut (black solid line),
 with $n\sigma_{\pi}>0$ cut (red dashed line), and
 $n\sigma_{p}<0$ cut (blue dotted line).}
 \label{m2_nsigcut}
\end{minipage}
\end{figure}
Fig.~\ref{TOFrPID} and Fig.~\ref{bigplot} show that TOF alone is
not able to separate between pions and kaons and TPC alone is not
able to separate pions, and protons at the \pT range roughly
between 2.0 and 4.0 GeV/$c$.  However, the dependence of $1/\beta$
on \pT from TOF and that of dE/dx on \pT are different in this \pT
range and therefore, by combining these two, we will be able to
extend our PID capability.  Fig.~\ref{nsigpi_m2} shows the hadron
distribution as a function of $n\sigma_\pi$ and mass square
($m^2$), with $m^2=p^{2}((t_{TOF}*c/l)^{2}-1)$, where p is the
momentum, $t_{TOF}$ is the time of flight, $c$ is the speed of
light in vacuum, and $l$ is the flight path length of the
particle. The pion, kaon and (anti-)proton peak can be clearly
seen in Fig. ~\ref{nsigpi_m2}. The solid line in
Fig.~\ref{m2_nsigcut} is the projection of Fig.~\ref{nsigpi_m2} to
the $m^{2}$ direction at $3<p_T<4$ GeV/$c$. At this $p_T$ range,
the pion and kaon bands are merged together, and cannot be clearly
separated from the (anti-)proton band with the TOF. However, if we
require $n\sigma_{\pi}>0$ and then plot the $m^{2}$ distribution
again (dashed line in Fig. \ref{m2_nsigcut}), the kaon and
(anti-)proton bands are greatly suppressed and a clear pion signal
is observed. This is due to the rather large difference between
$n\sigma_{\pi}^{\pi}$ and $n\sigma_{K}^{K}$ or
$n\sigma_{p(\bar{p})}^{p(\bar{p})}$ in this $p_T$ range, as shown
in Fig.~\ref{nsigchk}. Similarly, the pion and kaon bands are
suppressed significantly with respect to the (anti-)proton band
when $n\sigma_{p(\bar{p})}<0$ is applied. This helps us get a
cleaner (anti-)proton signal. Therefore, the combination of the
TPC and TOF enhances the particle identification at $2<p_T<4$
GeV/$c$ where neither TPC nor TOF can well separate the hadrons.

The invariant yields of pions, kaons and (anti-)protons were
calculated for 62.4 GeV Au+Au collisions. The results were shown
elsewhere in \cite{TOFHQ2004}.  The pions and (anti-)protons were
identified at $0.2 < p_T < \sim$ 5 GeV/$c$, and the kaons were
identified at $0.2 < p_T < \sim$ 3 GeV/$c$. The PID $p_T$ reach is due
to low statistics with the small acceptance of TOFr.

\subsection{Cross-check at low/intermediate $p_T$}
\begin{figure}[ht]
\begin{minipage}{0.45\textwidth}
{\includegraphics[width=0.9\textwidth] {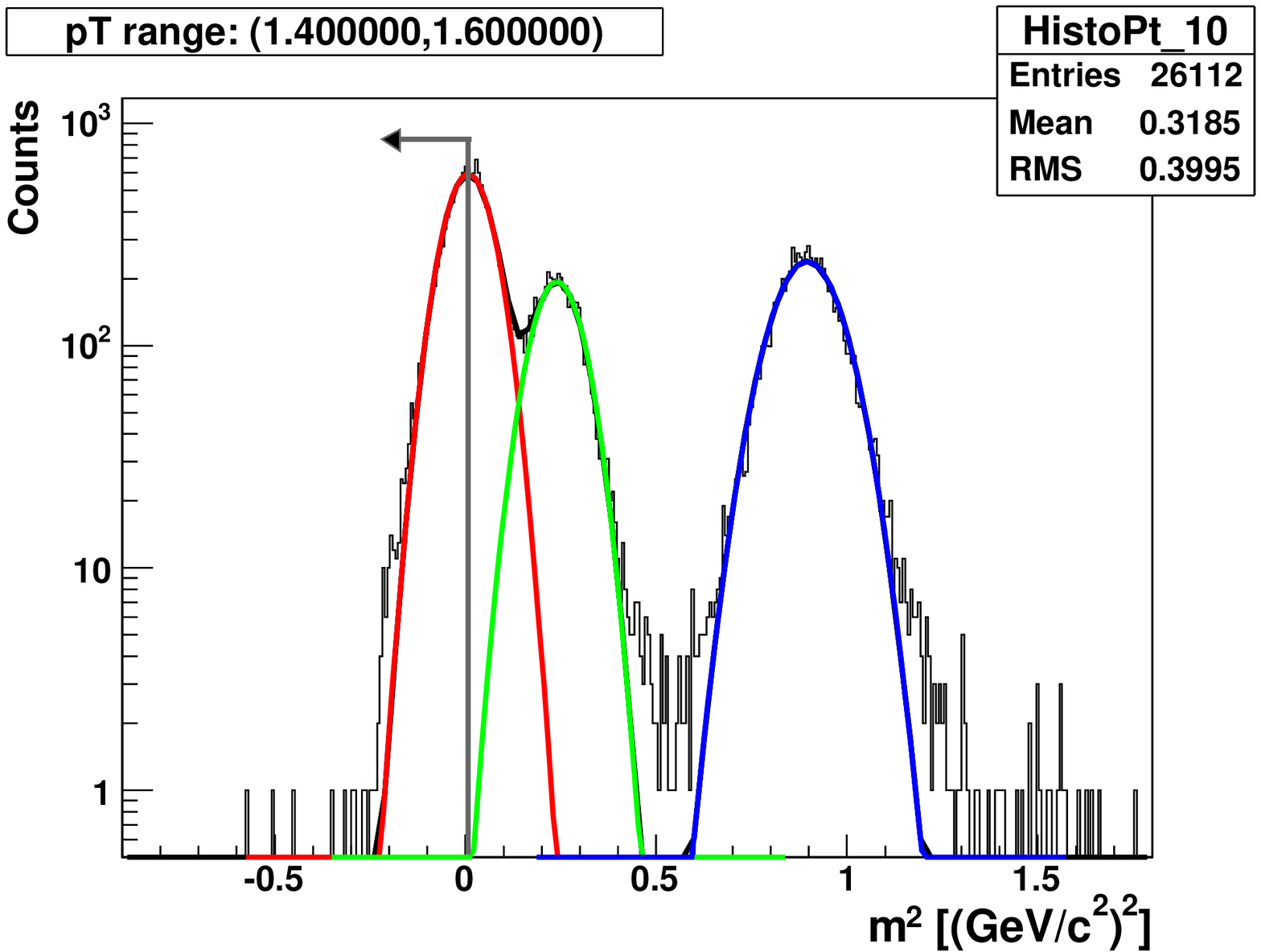}}
 \caption[]{$m^2$ distribution from TOFr for $1.4<p_T<1.6$ GeV/$c$.
Arrow shows the cut at the pion mass in order to obtain a clean pion
sample for dE/dx studies.}
 \label{pisample}
\end{minipage}
\begin{minipage}{0.1\textwidth}
\end{minipage}
\begin{minipage}{0.45\textwidth}
\begin{center}
{\includegraphics[width=1.0\textwidth] {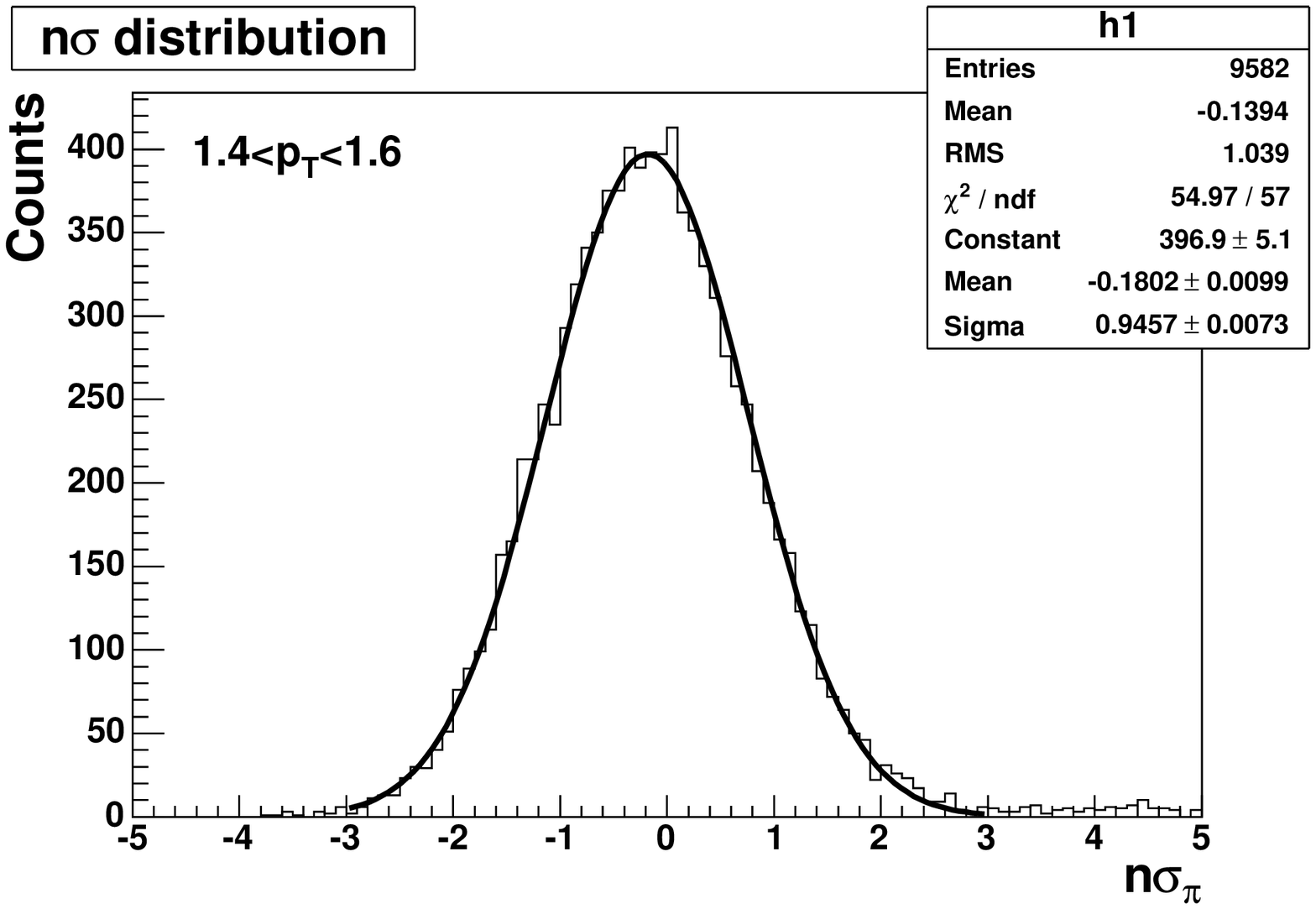}}
\end{center}
 \caption[]{$n\sigma_{\pi}$ distribution using the pion sample
 resulting from the cut described in previous figure. The solid lines
 are fits to a Gaussian function.}
 \label{nsigpara}
\end{minipage}
\end{figure}
\begin{table}{}
\caption{\label{fitchk}$n\sigma_{\pi}$ Gaussian fit parameters of
the pion and (anti-)proton sample obtained using $m^2$ cut. The
differences between $n\sigma_{\pi}^{p(\bar{p})}$ and
$n\sigma_{\pi}^{\pi}$ are also listed and compared with Bichsel
calculations. At $p_T \geq 3$ GeV/$c$,
$<n\sigma_{\pi}^{p(\bar{p})}$-$n\sigma_{\pi}^{\pi}>$ are
calculated via $h^{+}-h^{-}$.}{\centering
\begin{scriptsize}{\begin{tabular} {|c||c|c||c|c||c|c|} \hline
 $p_T$&\multicolumn{2}{c||}{$n\sigma_{\pi}^{\pi}$}&\multicolumn{2}{c||}{$n\sigma_{\pi}^{p(\bar{p})}$}
 &\multicolumn{2}{c|}{$<n\sigma_{\pi}^{p(\bar{p})}$-$n\sigma_{\pi}^{\pi}>$}\\
\cline{2-7}
 (GeV/$c$)&mean&sigma&mean&sigma&measurement&expectation\\
\hline \hline
0.9-1.0&-0.185$\pm$0.006&0.980$\pm$0.004&4.727$\pm$0.014&1.110$\pm$0.010&4.912$\pm$0.020&4.936\\
1.0-1.2&-0.171$\pm$0.005&0.977$\pm$0.004&3.176$\pm$0.021&1.171$\pm$0.015&3.347$\pm$0.026&3.421\\
1.2-1.4&-0.150$\pm$0.007&0.958$\pm$0.005&1.562$\pm$0.013&1.112$\pm$0.011&1.712$\pm$0.020&1.872\\
1.4-1.6&-0.180$\pm$0.010&0.946$\pm$0.007&0.464$\pm$0.015&1.073$\pm$0.012&0.645$\pm$0.025&0.768\\
1.6-1.8&-0.195$\pm$0.013&0.938$\pm$0.010&-0.322$\pm$0.019&1.021$\pm$0.014&-0.127$\pm$0.032&-0.036\\
1.8-2.0&-0.228$\pm$0.021&0.923$\pm$0.015&-0.834$\pm$0.023&0.989$\pm$0.016&-0.606$\pm$0.044&-0.616\\
2.0-2.5&-0.171$\pm$0.029&0.972$\pm$0.023&-1.458$\pm$0.025&1.034$\pm$0.019&-1.287$\pm$0.054&-1.174\\
2.5-3.0&&&-1.912$\pm$0.056&1.161$\pm$0.048&-1.91$\pm$0.01&-1.72\\
3.0-3.5&&&-2.387$\pm$0.113&1.111$\pm$0.121&-2.40$\pm$0.01&-2.06\\
3.5-4.0&&&&&-2.50$\pm$0.02&-2.27\\
4.0-4.5&&&&&-2.52$\pm$0.04&-2.35\\
4.5-5.0&&&&&-2.47$\pm$0.08&-2.44\\
5.0-6.0&&&&&-2.47$\pm$0.15&-2.47\\
6.0-7.0&&&&&-2.60$\pm$0.28&-2.47\\
 \hline
\end{tabular}
} \end{scriptsize}\par}
\end{table}
Since dE/dx plays a key role in our methods discussed above, dE/dx
calibration has to be carefully studied. Again, with PID capability at
overlapping $p_T$ range, we can use identified particles (pions,
protons etc.) from TOFr to cross-check the characteristics of
$n\sigma$($\pi$, K, $p(\bar{p})$) distribution at low/intermediate
$p_T$. Fig.~\ref{pisample} demonstrates an example of choosing a pion
sample by cutting on $m^{2}$. We can then plot the $n\sigma_{\pi}$
distribution of this sample and fit it with a Gaussian function as in
Fig.~\ref{nsigpara}. The mean and sigma of the Gaussian are listed in
Table~\ref{fitchk} as a function of $p_T$. The parameters (mean and
sigma) obtained from proton samples are also listed in Table
\ref{fitchk}. We note that the width of $n\sigma^{p}_{\pi}$ is
systematically larger than that of pions, partly due to the different
resolution from different track length resulting in dispersion of
$n\sigma^{p}_{\pi}$. The values of these parameters for pions and
(anti-)protons are almost independent of $p_T$ and consistent with
that obtained by pure dE/dx method (see Fig.~\ref{rdEdxpid}).  The
last two columns in Table~\ref{fitchk} show the measured
$<n\sigma_{\pi}^{p(\bar{p})}$-$n\sigma_{\pi}^{\pi}>$ compared to
Bichsel function calculation. The experimental results agree well with
the expectations within errors.

\subsection{Contamination in pion identification}
\begin{figure}[ht]
\begin{center}
{\includegraphics[width=0.7\textwidth] {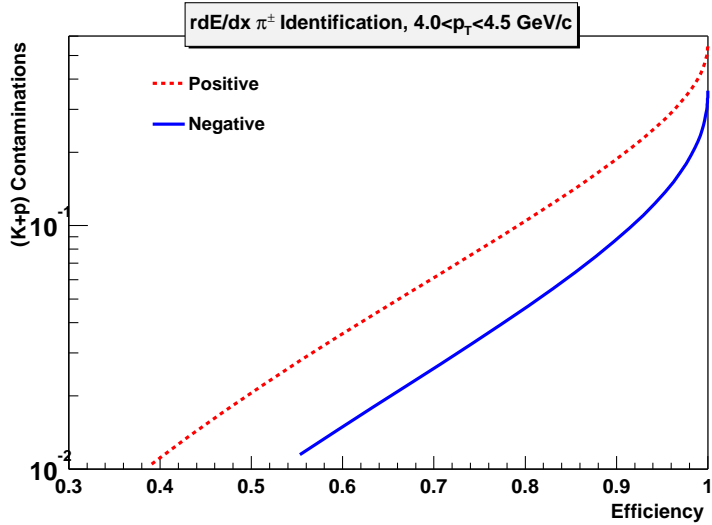}}
\end{center}
 \caption[]{Contamination from Kaons+(anti-)protons to pions with a
dE/dx selection window vs pion efficiency from the same selection at
$4 < p_T <4.5$ GeV/$c$. Dashed line is for $\pi^+$ and solid line is
from $\pi^-$.}
 \label{picontam}
\end{figure}
Good particle-by-particle identification is crucial for analyses
such as collective flow and fluctuation which provide important
physics at RHIC. Not only can pion yields be obtained by
statistically fitting of the histogram, pion identification can
also be obtained track by track using a dE/dx cut on the
$n\sigma_{\pi}$. Fig.~\ref{picontam} illustrates the contamination
to pions in percentage from kaons and (anti-)protons when we
select $n\sigma_{\pi}$ greater than a given value (threshold). We
can achieve $> 95\%$ pion purity at $50\%$ pion efficiency when
requiring $n\sigma_{\pi}>0$. As have been shown previously,
$n\sigma_{\pi}^{\pi}$ is actually not centered at zero due to
insufficient calibration. However, this can be easily corrected
for. If one can tolerate a pion contamination of $10\%$, then the
efficiency can be as high as $\sim 80\%$. These represent the
worst-case scenario since at higher $p_T$ the baryon enhancement
is less and the contamination should be smaller.

\subsection{Contamination in proton identification}
\begin{figure}[ht]
\begin{center}
{\includegraphics[width=0.7\textwidth] {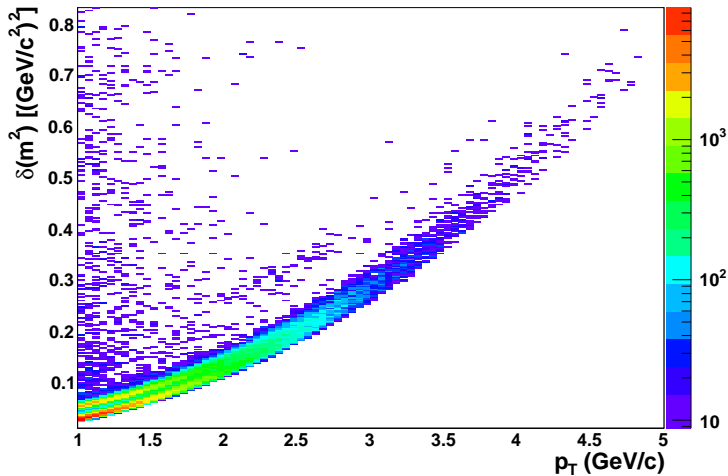}}
\end{center}
 \caption[]{$m^2$ resolution as a function of $p_T$.}
 \label{m2resolution}
\end{figure}

The contamination in (anti-)proton identification is a little more
complicated. As shown in Fig.~\ref{bigplot}, the separation of
dE/dx between protons and kaons is $\sim 1\sigma$ at
intermediate/high $p_T$. Therefore the purity of protons at
$p_T<4$ GeV/$c$ is mainly driven by the $m^{2}$ resolution of
TOFr. The $m^{2}$ resolution is formulated by:

$$\delta{m^2} = \frac{\delta{p^2}}{\beta^{2}\gamma^{2}} \oplus
\frac{p^2}{\beta^{2}}\frac{2\delta{t}}{t} \oplus
\frac{p^2}{\beta^{2}}\frac{2\delta{L}}{L},$$

where $p$ is the momentum, $t$ is the time of flight and $L$ is the
track length of the particle. Fig.~\ref{m2resolution} depicts the
$m^2$ resolution for all charged particles as a function of $p_T$, at
$-0.5 \leq\eta\leq 0$. The momentum resolution is taken from
\cite{TPC}. $\delta{t}$ and $\delta{L}$ are chosen to be 110ps and
0.5cm, respectively. At intermediate/high $p_T$, the $m^2$ resolution
increases approximately quadratic with momentum.  If we require $m^{2}
\geq 0.88$ (mass square of $p(\bar{p})$) and $n\sigma_{p(\bar{p})} <
0$, which means an (anti-)proton efficiency of $25\%$, the
contamination from pions and kaons is $\sim 10\%$ at $4 < p_T < 5$
GeV/$c$. We have used the invariant yields of $\pi$/K/p from
\cite{TOFHQ2004}\cite{K0s} in this estimation.

At higher $p_T$ ($>5$ GeV/$c$), TOFr is not effective in identifying
(anti-)protons anymore. The contamination to (anti-)proton
identification from kaons is mainly determined by the
${}^{>}_{\sim}1\sigma$ separation between them.  Therefore, at $50\%$
efficiency ($n\sigma_{p(\bar{p})} < 0$), kaon contamination is about
20\% assuming kaon and proton yields are equal.  Again we can use
different methods to evaluate the contamination and efficiency as
described in previous section.

\begin{figure}
\begin{minipage}{0.45\textwidth}
{\includegraphics[width=0.9\textwidth]
{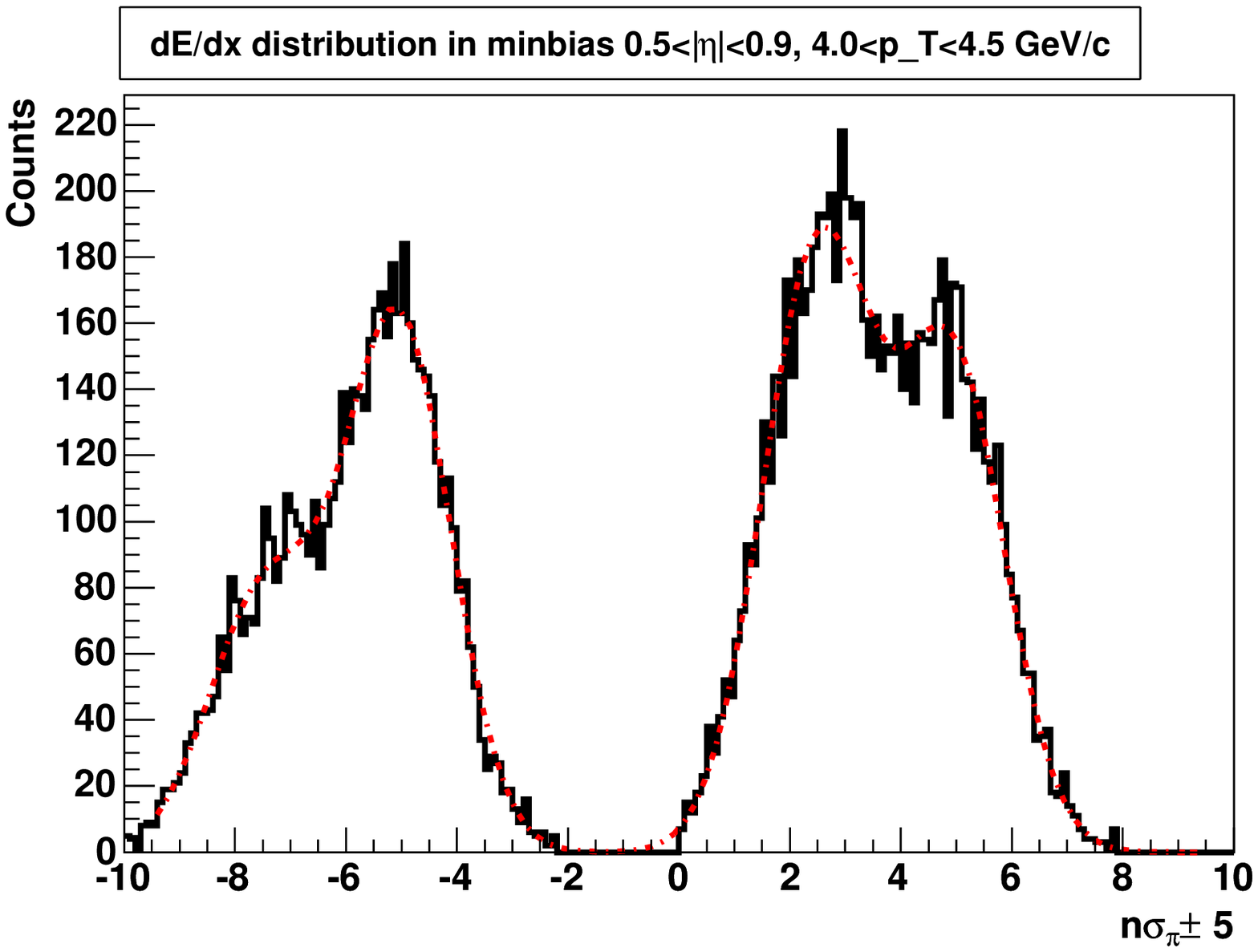}}
 \caption[]{\small{dE/dx distribution normalized by pion dE/dx and offset by
$\pm5$ for positive and negative charge at $4<p_T<4.5$ GeV/$c$ and
$0.5 < |\eta| < 0.9$, respectively. The distribution is from
minimum-bias Au+Au collisions at $\sqrt{s_{_{NN}}}=62.4$ GeV.}}
 \label{largereta}
\end{minipage}
\begin{minipage}{0.1\textwidth}
\end{minipage}
\begin{minipage}{0.45\textwidth}
{\includegraphics[width=0.9\textwidth]
{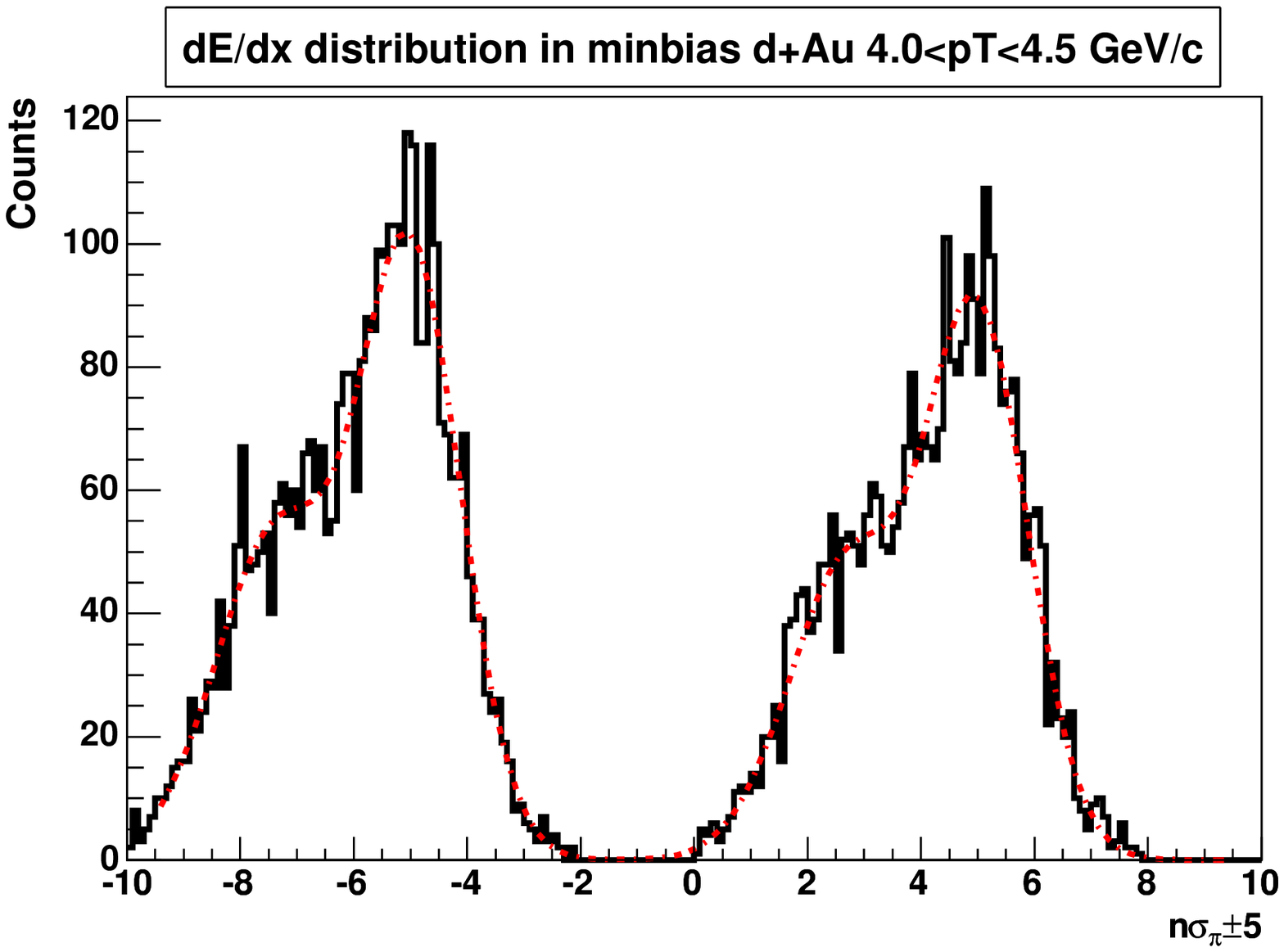}}
 \caption[]{\small{dE/dx distribution normalized by pion dE/dx ($4<p_T<4.5$ GeV/$c$) and offset by
$\pm5$ for inclusive hadrons at $0.5 < \eta < 1.0$ and $-1.0 <
\eta < -0.5$, respectively. The distribution is from minimum-bias
d+Au collisions at $\sqrt{s_{_{NN}}}=200$ GeV.}}
 \label{dAu}
\end{minipage}
\end{figure}
At STAR, the resolution of dE/dx measurement depends on many
factors, including magnetic field setting, event multiplicity,
beam luminosity, track length and drift distance. It also depends
on the number of hits in TPC used to calculate the ionization
energy loss for a given track. Excellent readout electronics
system (pulse shape control, fine linearity and large dynamic
range, etc.) and careful offline calibration are important to
achieve high-quality dE/dx measurement. Particle identification
may have difference efficiency and purity, depending on the
experimental setting and cuts used in analysis. Basically, dE/dx
resolution is improved with longer track length, shorter drift
distance, stronger magnetic field, lower multiplicity and beam
luminosity, as well as more hits in TPC used for track fitting.
Fig.~\ref{largereta} and~\ref{dAu} shows pion identification at
$0.5 < |\eta| < 0.9$ in Au+Au collisions at $\sqrt{s_{_{NN}}}=200$
GeV and $1.0>|\eta|>0.5$ in d+Au collisions at $4 \leq p_T \leq
4.5$ GeV/$c$, respectively. Due to longer track and shorter drift
distance for particles produced at higher $|\eta|$, the dE/dx
resolution gets better. Thus the separation between pions and
Kaons or(anti-)protons in Fig.~\ref{largereta} and Fig.~\ref{dAu}
are larger than that in Fig.~\ref{rdEdxpid}, and better
identification can be achieved. From the relative heights of pion
and proton peaks, it is also obvious that the particle
compositions in Au+Au at \sNN=62.4 GeV and d+Au at \sNN=200 GeV
are different.

\section{Electron Identification at low/intermediate $p_T$}
\begin{figure}[ht]
\begin{center}
{\includegraphics[width=0.7\textwidth] {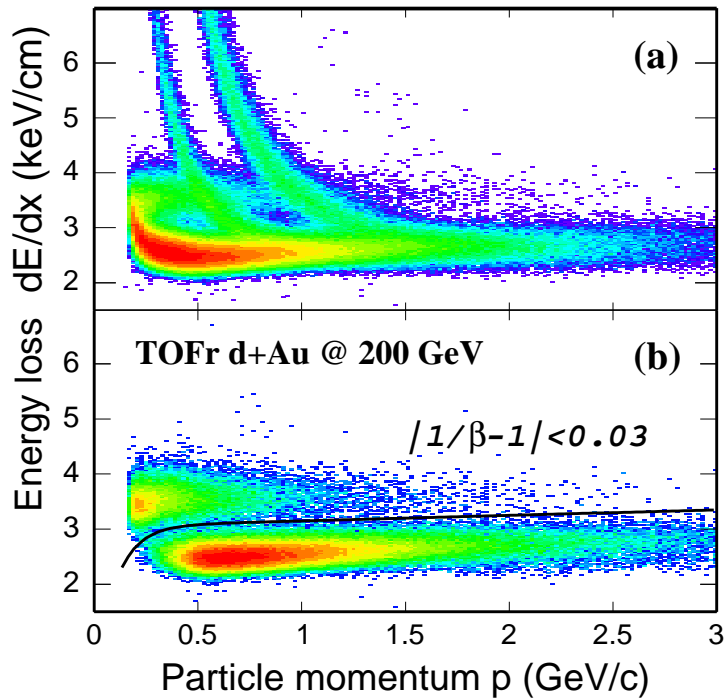}}
\end{center}
 \caption[]{dE/dx in the TPC vs. particle momentum ($p$) without
 (upper panel) and with (lower panel) TOFr velocity cut of
 $|1/\beta-1|<0.03$.}
 \label{ePID}
\end{figure}
In addition to its hadron PID capability, TOFr detector can be
used to identify electrons by combining it with dE/dx from TPC. We
were not able to measure dilepton with the prototype MRPC TOF tray
installed in STAR due to small TOF acceptance.  However, single
electron spectra are sensitive to charm production and future
azimuthal $2\pi$ coverage of TOF will enable us to do dilepton
physics\cite{TOFopencharm}. In this section, we will discuss the
electron identification method, the hadron rejection power at low
\pT and rejection of the photonic background from $\gamma$
conversions and Dalitz decays~\cite{xinthesis}.

The top panel of Fig. ~\ref{ePID} shows the 2-D scatter plot of dE/dx
as a function of the particle momentum ($p$) for charged particles
with TOFr matched hits from $d$+Au collisions. The bottom panel shows
that slow hadrons are eliminated and the electron band is well
separated from the hadron band with a particle velocity ($\beta$)
requirement at $|1/\beta-1|<0.03$. Electrons can then be selected with
the following cut (\ref{tofCut}):
\begin{equation}
dE/dx(p)>2.4+0.65\times(1-e^{-(p-0.15)/0.1})+0.1\times p\label{tofCut}
\end{equation}
where $p$ is in GeV/$c$ and dE/dx is in keV/cm. With
the combination of dE/dx from TPC and $\beta$ from TOFr, electrons
can be identified above $p \sim 0.15$ GeV/$c$, while the high
$p_T$ reach is limited by the statistics in this analysis.

\subsection{contamination for electron identification}
\begin{figure}
\begin{center}
\includegraphics[width=0.45\textwidth]{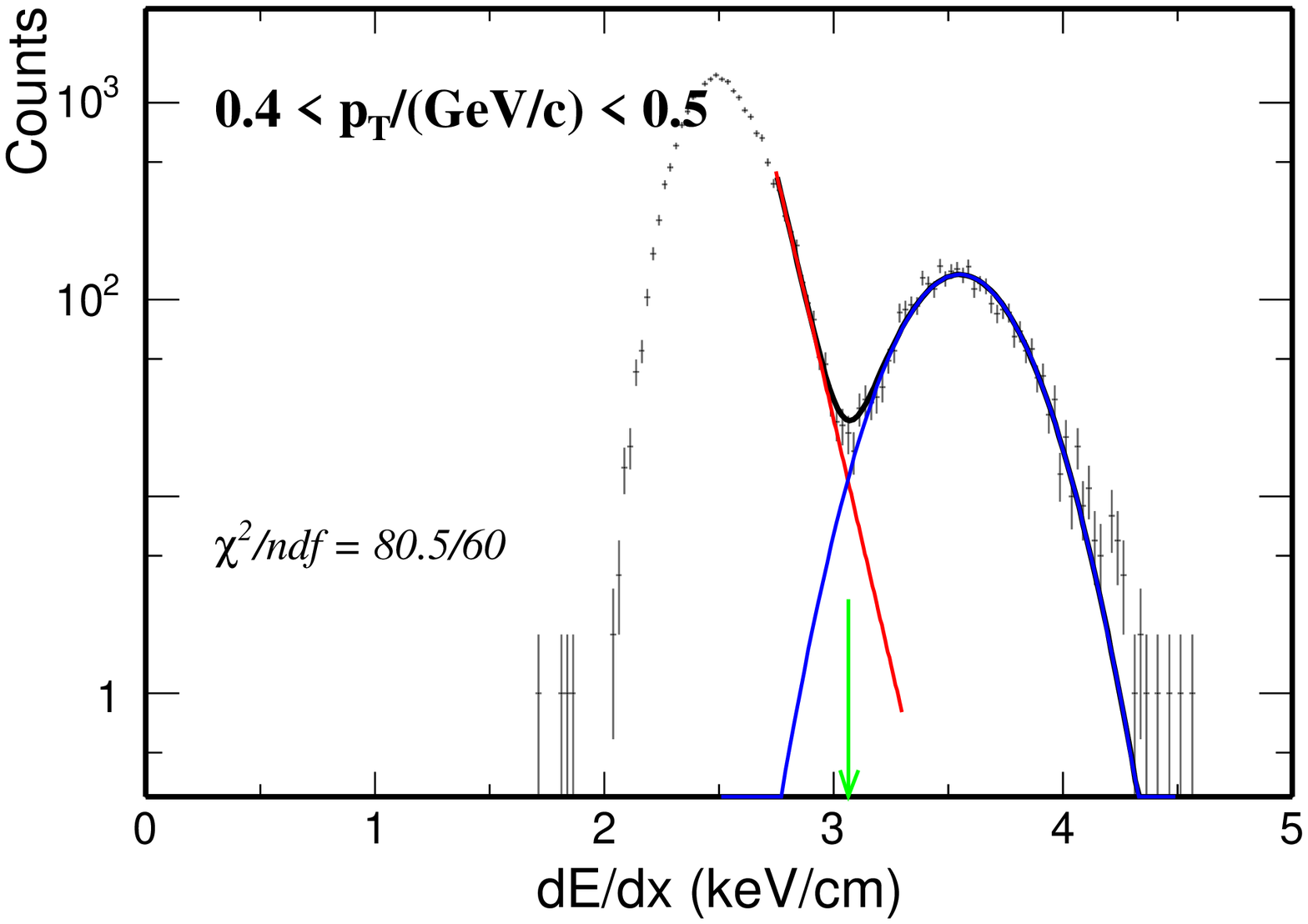}
\includegraphics[width=0.45\textwidth]{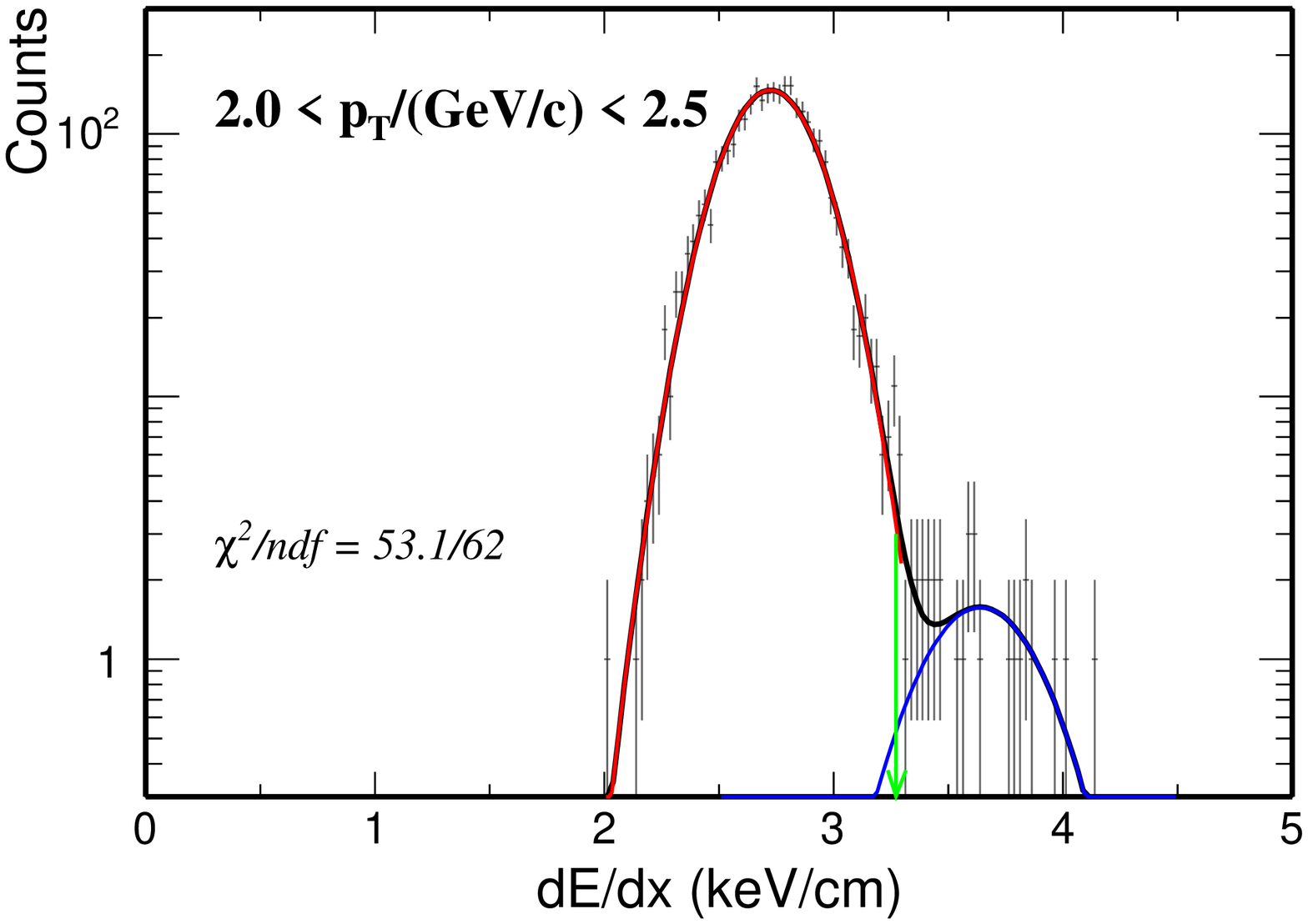}
\end{center}
\caption[]{dE/dx distribution with TOFr velocity cut in 2 $p_T$
bins. The arrows denote the cut from Eq.(~\ref{tofCut}). The curve
in left plot ($0.4<p_T<0.5$GeV/c) shows a Gaussian plus
exponential fit to the dE/dx distribution, while that in right
plot ($2.0<p_T<2.5$GeV/c) is a 2-Gaussian fit.} \label{dEdxFitTof}
\end{figure}

\begin{figure}[ht]
\begin{center}
{\includegraphics[width=0.7\textwidth] {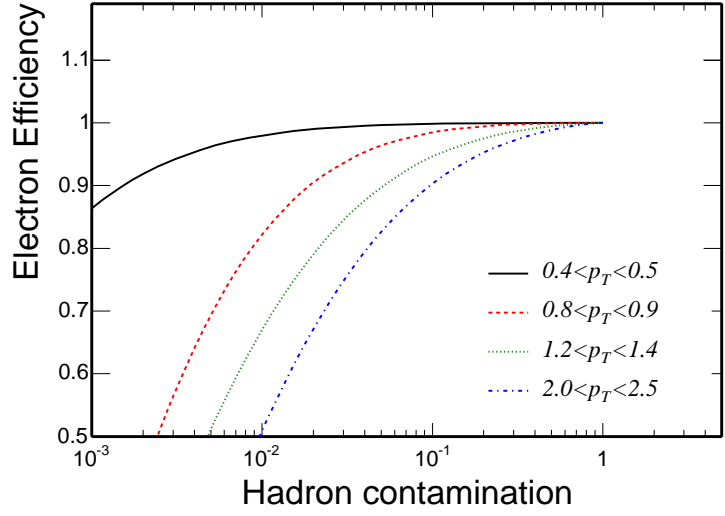}}
\end{center}
 \caption[]{The contamination from hadrons to electron
 identification as a function of electron efficiency in different
 $p_T$ range, with TOFr velocity cut.}
 \label{hadronCom}
\end{figure}

\begin{figure}[ht]
\begin{center}
{\includegraphics[width=0.7\textwidth] {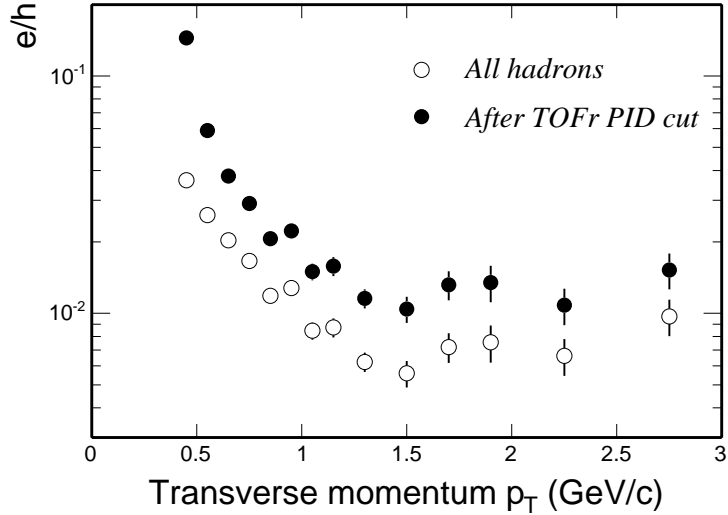}}
\end{center}
 \caption[]{Electron/hadron ratio vs. transverse momentum ($p_T$),
  with and without TOFr velocity cut.}
 \label{e2H}
\end{figure}

Hadron contamination to electron identification with TOFr velocity
cut was studied from the dE/dx distribution in each $p_T$ bin. A
Gaussian function with an exponential tail is used in the fit. At
$p_T$ $\simeq 2-3$ GeV/$c$, 2-Gaussian fit is also performed and
shows little difference from the Gaussian plus exponential fit.
Fig.~\ref{dEdxFitTof} shows the results in two $p_T$ bins from
d+Au collisions. The arrows denote the cut from Eq.
(~\ref{tofCut}). Hadron contamination ratio is estimated from
these fits, and shown in Fig.~\ref{hadronCom}. The electron
efficiency is obtained by varying the cut in Eq. (~\ref{tofCut}).
At low $p_T$, high electron efficiency and low hadron
contamination can be achieved. Even at intermediate $p_T$, the
contamination can be reduced to a low level ($\sim 1\%$) if a
lower electron efficiency ($\sim 50\%$) is chosen. Fig.~\ref{e2H}
shows electron to hadron ratio vs. $p_T$ before and after TOFr
velocity selection. The hadron rejection power can be evaluated by
(hadron contamination)$\times$($e/h$)/(electron efficiency) for a
specific condition. It was estimated to be at $10^{-5}$ level at
$p_T<1.0$ GeV/$c$. At higher $p_{T}$, additional hadron rejection
from electromagnetic calorimeter in similar fiducial coverage can
help reach the same rejection power~\cite{EMC}.


\subsection{Single electrons from Au+Au collisions at \sNN = 62.4 GeV}

The purpose of the data analysis of d+Au and p+p collisions is to
set up the baseline for heavy ion collisions. Since the large data
sample of Au+Au collisions at \sNN = 200 GeV is not available yet,
the smaller sample of Au+Au collisions at \sNN = 62.4 GeV has been
analyzed to understand the necessary techniques for Au+Au
collisions. There is a significant increase of multiplicities in
Au+Au collisions as compared to that in p+p and d+Au collisions,
and a consequent decrease of track quality. In this section,
we discuss the electron measurements from TOF and dE/dx
method and the photonic background estimation. Since the charm
yield is low compared to photonic contributions at \sNN =
62.4 GeV, any excess of electrons above the photonic background
with acceptable errors is not expected. This provides a testing
ground to evaluate the quality of background reconstruction.

STAR has accumulated $\sim15$ million events in a relatively short run
with Au+Au collisions at \sNN = 62.4 GeV. With the minimum bias
trigger ($0-80\%$) and vertex z position selection, the events used in
this analysis are $\sim6.4$ million.  The resolution for TOF detectors
is $\sim110$ ps for TOF system, with $\sim55$ ps start timing
resolution included. The hadron PID capability was reported
in~\cite{TOFHQ2004}. As in d+Au and p+p collisions, electrons can be
identified by combining TOF and dE/dx in the TPC. Electrons were
selected to originate from the primary interaction vertex according to
the criteria shown in Table ~\ref{eTrackAuAu}.

\begin{table}[hbt]
\caption{Electron selection criteria in Au+Au}
\label{eTrackAuAu}\vskip 0.1 in \centering\begin{tabular}{c|c}
\hline \hline Method & TOF+dE/dx       \\ \hline $|VertexZ|<$ & 30
cm
\\ \hline
primary track ?      &   Yes             \\
nFitPts $>$    &  25               \\
ndEdxPts $>$   & 15                \\
rapidity             & (-1.0, 0)         \\
$\chi^{2}/ndf$       & (0., 3.0)         \\
$\beta$ from TOF     & $|1/\beta-1|<0.03$\\
TOFr hit quality     & $30<ADC<300$      \\
                     & $-2.7<z_{local}$/cm$<3.4$ \\
                     & $|y_{local}-y_{C}|<1.9$ cm \\
TOFp hit quality     & $th_1<ADC<th_2$    \\
                     & $2.0<z_{local}$/cm$<18.0$ \\
                     & $0.4<y_{local}$/cm$<3.2$ \\
     \hline \hline
\end{tabular}
\end{table}

\begin{figure}
\begin{center}\includegraphics[width=0.7\textwidth]{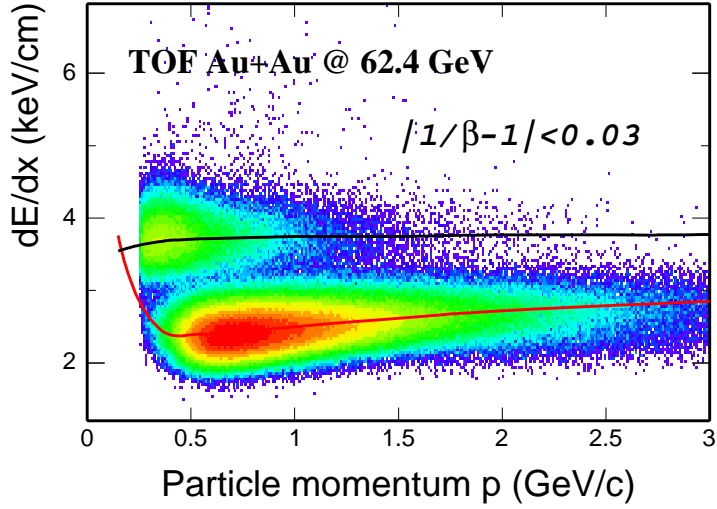}\end{center}
\caption[Electron PID in Au+Au 62.4 GeV]{dE/dx vs. particle
momentum after a TOF $\beta$ cut ($|1/\beta-1|<0.03$)}
\label{ePIDAuAu} \end{figure}

Fig.~\ref{ePIDAuAu} shows the dE/dx vs. particle momentum after a
$\beta$ cut from TOF. The electron dE/dx band can be separated from
that of hadrons. The dE/dx resolution in Au+Au collisions
decreases when compared to that in d+Au and p+p due to much higher
multiplicities and a multiplicity dependence of dE/dx calibration
which is not currently implemented. We fit the dE/dx distribution
around electron peak with both two-gaussian function and
exponential+gaussian function to extract the electron raw yields
in each \pT bin.

\begin{figure}
\begin{center}\includegraphics[width=0.7\textwidth]{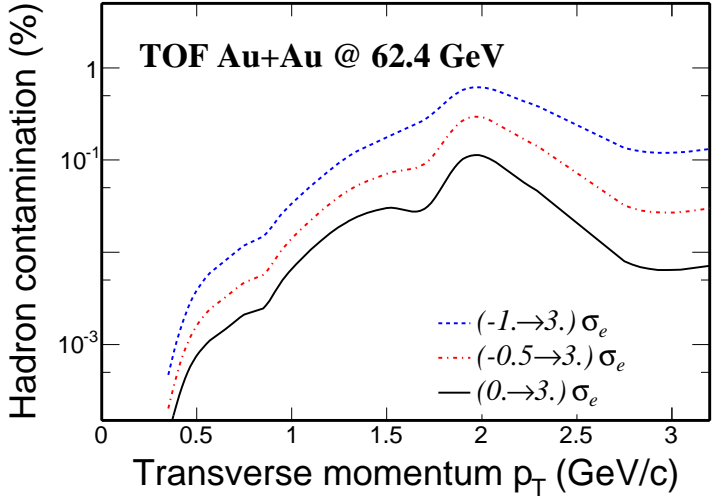}\end{center}
\caption[]{Hadron contamination fractions for different electron
  selections in Au+Au 62.4 GeV collisions.} \label{hadComAuAu} \end{figure}

Hadron contamination becomes larger at $p_T>1.5$ GeV/$c$ if we
select electrons with the same efficiency as in p+p and d+Au
collisions. Fig.~\ref{hadComAuAu} shows the hadron contamination
fractions under different electron selections. The selection
$\sigma_{e}>0$ with electron efficiency of about $50\%$ is
necessary to have the contaminations less than 10\% at $p_T>1.5$
GeV/$c$. At lower \pT, the identification is not affected as much
by the high multiplicity since most of the hadrons are rejected by
time-of-flight selection.

\subsection{Estimate of electrons from photonic source in Au+Au}
\begin{table}[hbt]
\caption{Partner candidate selection criteria in Au+Au}
\label{ePartnerAuAu}\vskip 0.1 in \centering\begin{tabular}{c|c}
\hline \hline charge &  opposite to tagged track \\
primary/global ?     &   global             \\
nFitPts $>$  &  15               \\
nFitPts/nMax $>$     &  0.52          \\
$\chi^{2}/ndf$       & (0., 3.0)         \\
$\sigma_{e}$         & (-1., 3.0)$^\dagger$ \\ \hline $dca$ of
$e^{+}$,$e^{-}$ & (0.0, 3.0) cm \\
     \hline \hline
\end{tabular} \\
$^\dagger$ several different $\sigma_{e}$ cuts have been studied.
\end{table}

\begin{figure}
\begin{center}\includegraphics[width=0.7\textwidth]{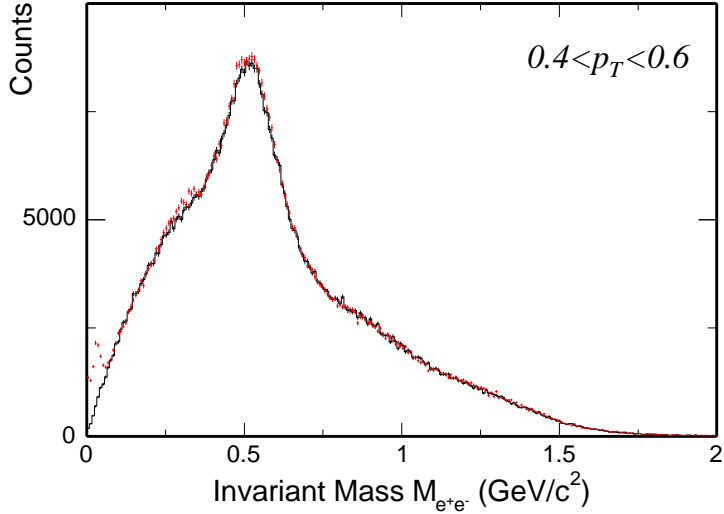}\end{center}
\caption[]{Main photonic background reconstruction in Au+Au
collisions. The crosses depict the real electron pair candidate
invariant mass distributions and the histograms represent the
combinatorial backgrounds. The photon conversion peak is clear in
$\sim 0$ mass region, while the $\pi^{0}$ Dalitz is hard to see.}
\label{eAuAubkgd} \end{figure}

Photon conversions $\gamma\rightarrow e^{+}e^{-}$ and
$\pi^0\rightarrow\gamma e^{+}e^{-}$ Dalitz decays are the dominant
photonic sources of electron background.  Since only one TOF tray
(120th of the proposed TOF) is available, electron identification
of both $e^{+}e^{-}$ daughters from conversion or Dalitz decays is
not possible at this momentum. To measure the background photonic
electron spectra, the invariant mass and opening angle of the
$e^+e^-$ pairs were constructed from an electron (positron) in
TOFr and every other positron (electron) candidate reconstructed
in the TPC~\cite{johnson}. A secondary vertex at the conversion
point was not required.  In Au+Au collisions, due to large
multiplicity, the partner track reconstruction will lead to a
large combinatorial background due to hadron contamination without
an additional PID from TOF. Even with a stringent dE/dx selection,
this combinatorial background is still significant. In the future,
dilepton studies would be possible with $2\pi$ coverage of TOF for
selecting clean $e^{+}e^{-}$ pairs. Fig.~\ref{eAuAubkgd} shows the
electron pair candidate invariant mass distribution. The tagged
electron was selected from TOF with the cuts shown in Table
~\ref{eTrackAuAu} and additional $0<n\sigma_{e}<3$. The partner
track candidate was selected according to Table
~\ref{ePartnerAuAu}.

 The combinatorial background
distribution was provided by rotating the partner track momentum
$\overrightarrow{p}\rightarrow -\overrightarrow{p}$, and
normalized to the distribution of the real electron pair
candidates in the region $0.8<M_{e^+e^-}$/$<2.0$ (GeV/$c^2$) in
the invariant mass spectrum.  Fig.~\ref{eAuAubkgd} shows the
results in $0.4<p_T<0.6$ GeV/$c$ where \pT is the \pT of tagged
electrons. The plot shows that the combinatorial background was
very well reproduced. The peak from photon conversion in detector
material is clearly visible near zero mass region, and its offset
from zero is due to the opening angle resolution~\cite{johnson} in
TPC tracking. The $\pi^0$ Dalitz contribution is not visible in
this case, possibly because the Dalitz contribution is much
smaller and distribution is much broader compared to conversion
processes. We subtracted the combinatorial background from the
real distribution, and integrated the remaining distribution from
$0-0.15$ GeV/c$^{2}$ to get the reconstructed main photonic
background raw yield. In this case, we assumed that both photon
conversion and $\pi^{0}$ Dalitz decays were reconstructed.

\begin{figure}
\begin{minipage}{0.45\textwidth}
\begin{center}{\includegraphics[width=0.9\textwidth]{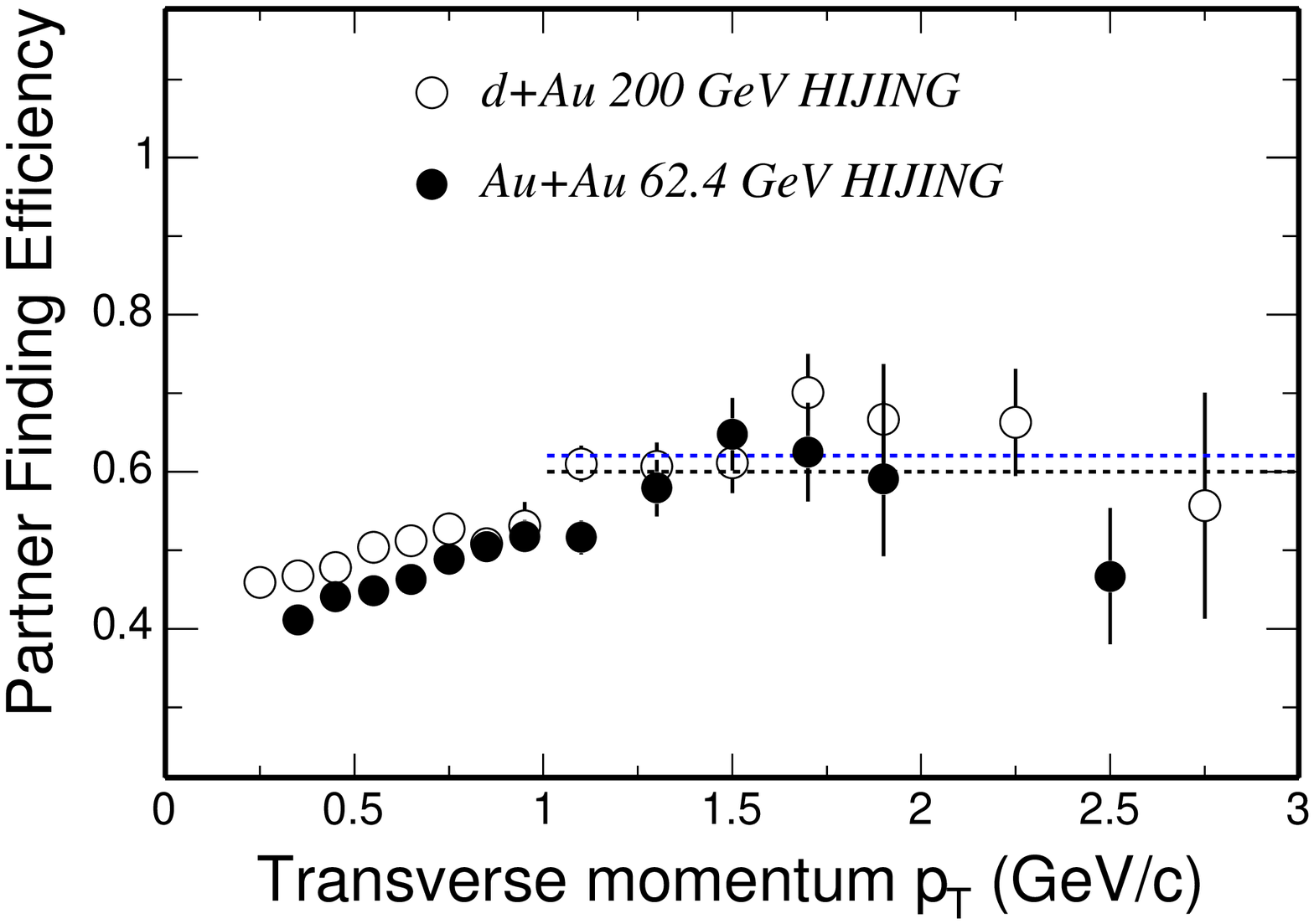}}\end{center}
\caption[]{Photonic background reconstruction efficiency from
Au+Au 62.4 GeV HIJING simulations. Also shown on the plot is that
from d+Au 200 GeV HIJING simulations.} \label{bkgdeff62}
\end{minipage}
\begin{minipage}{0.1\textwidth}
\end{minipage}
\begin{minipage}{0.45\textwidth}
\begin{center}{\includegraphics[width=0.9\textwidth]{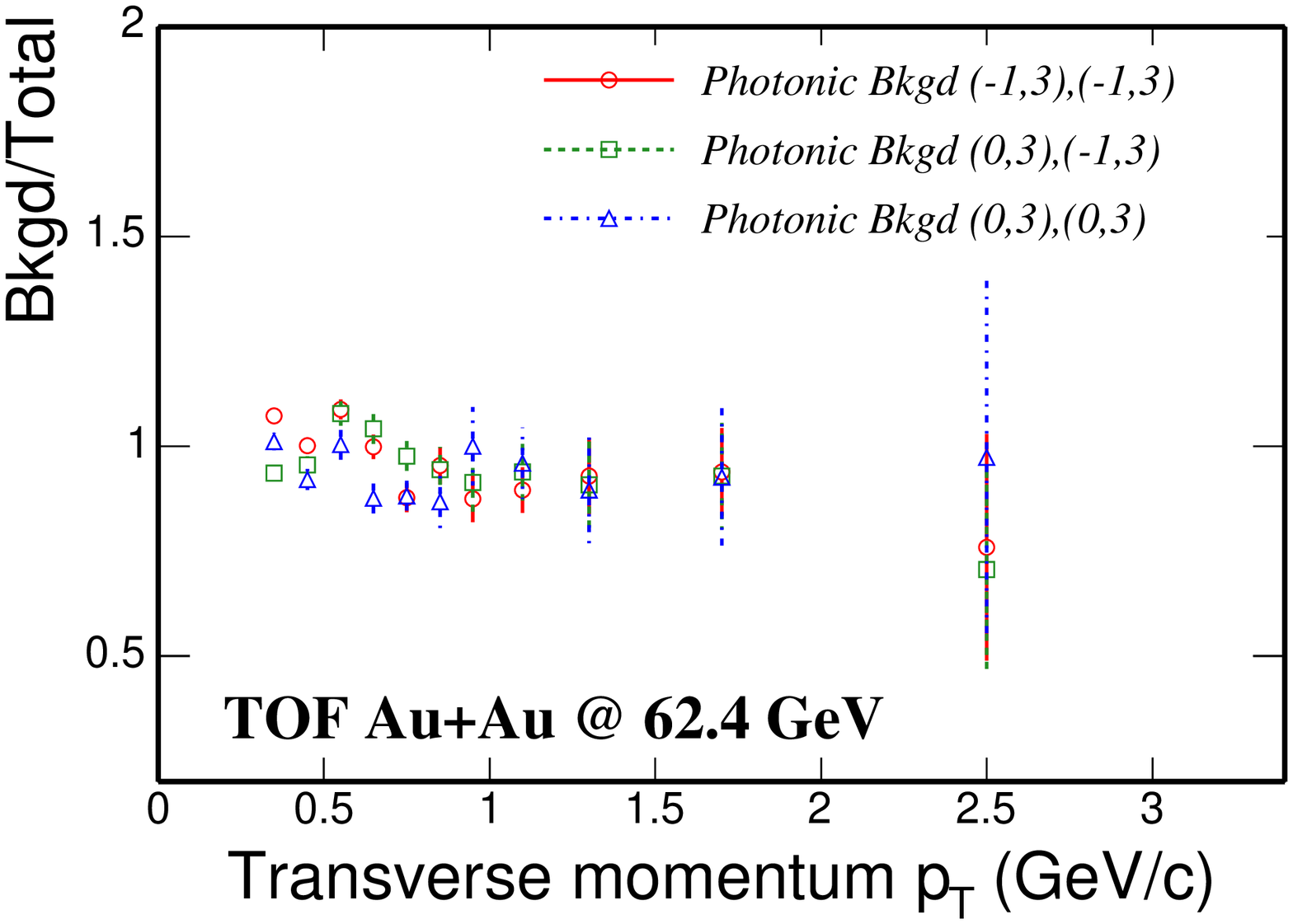}}\end{center}
\caption[]{The reconstruction efficiency-corrected photonic
background over the raw inclusive electron yield under different
electron/positron track selections. The numbers in the brackets on
the plot show the $\sigma_{e}$ cut for the tagged track and the
partner track, respectively.} \label{rawCom62}
\end{minipage}
\end{figure}

The background raw yield need to be corrected for the
reconstruction efficiency, as we did it for d+Au and p+p
collisions~\cite{TOFopencharm}. This efficiency was calculated
from Au+Au 62.4 GeV HIJING events with full detector MC
simulations. After $|V_{Z}|<30$ cm cut, $\sim53$ K events were
used in the calculation. We took all TPC electron tracks without
any dE/dx and TOF hit cut to improve the statistics. The procedure
is the same as we did in p+p and d+Au
collisions~\cite{TOFopencharm}. Fig.~\ref{bkgdeff62} shows the
background reconstruction efficiency in Au+Au 62.4 GeV compared
with d+Au results. Due to the higher multiplicities, the
reconstruction efficiency is relatively lower in Au+Au than that
in d+Au.

This efficiency was used to correct for the photonic background
raw yield obtained from above. In addition, a $\sim5\%$ fraction
of other photonic background (from d+Au results) and the
$n\sigma_{e}$ selection efficiency were also included. The
corrected background spectrum is compared with the inclusive
spectrum as shown Fig.~\ref{rawCom62}. The reconstructed
background matches the total inclusive electron spectrum. This is
expected since the charm yield is more than an order of magnitude
lower than the inclusive electron yield at 62.4 GeV~\cite{xinthesis}.
This also demonstrates that we can reconstruct fully the photonic
background from $\gamma$ conversions and Dalitz decays.

The performance of the electron identification and background
studies for Au+Au system makes us confident on extracting the charm
signal at $2-3$ GeV/c from the coming 30 M minimum bias 200 GeV Au+Au
data. With the STAR upgrade of inner tracker --Heavy Flavor Tracker
(HFT), we will be able to reject photon conversions by requiring
additional hits in HFT, to reject electrons from charm semileptonic
decays by its displaced secondary vertex, and to further
reconstruct/reject the low \pT partner of the electrons from the
Dalitz decays using inner tracker only.  Electron identification at
$p_T>1$ GeV/$c$ can be improved by combining dE/dx from TPC, velocity
from TOF and energy from electromagnetic calorimeter. Detailed studies
are underway and are beyond the scope of this article.

\section{Conclusion}
In summary, we have developed a technique to extend the particle
identification up to high $p_T$ at STAR. By combining information from
the TPC and TOF, we can measure pion, kaon and (anti-)proton in the
intermediate/high $p_T$ range and electron in low/intermediate $p_T$
(high $p_T$ limited by statistics).  Preliminary spectra obtained in
Au+Au collisions at \sNN=62.4 GeV shows reliable particle
identification to $p_T \sim 7-8$ GeV/$c$ for pions and (anti-)protons,
and $\sim3$ GeV/$c$ for kaons, respectively. The electrons were
identified at $0.15 < p_T < 4$ GeV/$c$, in d+Au, p+p collisions at
\sNN=200 GeV and Au+Au collisions at \sNN=62.4 GeV. A purity of over
$95\%$ for pion identification and $\sim 70-90\%$ for proton
identification was obtained with this method at intermediate/high
$p_T$. We achieved a hadron rejection power at $10^{-5}$ level for the
electron identification at low $p_T$.

\section{Acknowledgement}
We thank the STAR Collaboration, the RHIC Operations Group and RCF at
BNL, and the NERSC Center at LBNL for their support. This work was
supported in part by the HENP Divisions of the Office of Science of
the U.S. DOE; the Ministry of Education and the NNSFC of China.

\label{}


\begin{thebibliography}{00}




\bibitem{STARwhitepaper}STAR whitepaper:''Experimental and Theoretical
Challenges in the Search for the Quark Gluon Plasma: The STAR
Collaboration's Critical Assessment of the Evidence from RHIC
Collisions'', nucl-ex/0501009.
\bibitem{STARoverview}K.H. Ackermann {\it et al.}, \Journal{\NIMA}{499}{624}{2003}.
\bibitem{TPC}M. Anderson {\it et al.}, \Journal{\NIMA}{499}{659}{2003}.
\bibitem{TPCFEE}M. Anderson {\it et al.}, \Journal{\NIMA}{499}{679}{2003}.
\bibitem{TOF}The STAR TOF Collaboration, {\it Proposal for a Large
Area Time of Flight System for STAR}.
\bibitem{TOFcronin}J. Adams {\it et al. (STAR)}, nucl-ex/0309012; Lijuan Ruan,
Ph.D. thesis, University of Science and Technology of China,
nucl-ex/0503018.
\bibitem{TOFHQ2004}M. Shao {\it (STAR)}, Hot Quarks 2004, July 18-24,
2004, Taos Valley, New Mexico, USA; J. Phys. G: Nucl. Part. Phys.
31 (2005) S85-S92; L. Ruan {\it (STAR)}, SQM 2004, Sep. 15-20,
2004, Cape Town, South Africa.
\bibitem{TOFopencharm}J. Adams {\it et al. (STAR)},
\Journal{\PRL}{94}{2005}{062301}; nucl-ex/0407006.
\bibitem{MRPC}E. Cerron Zeballos{\it et al.}, \Journal{\NIMA}{374}{132}{1996}.
\bibitem{TOFr}B. Bonner {\it et al.}, \Journal{\NIMA}{508}{181}{2003};
M. Shao {\it et al.}, \Journal{\NIMA}{492}{344}{2002}.
\bibitem{TOFp}W.J. Llope {\it et al.}, \Journal{\NIMA}{522}{252}{2004}.
\bibitem{rdEdx}Z. Xu {\it (STAR)}, Division of Particles and Fields 2004, Aug. 26-31,
2004, Riverside, CA, USA; nucl-ex/0411001.
\bibitem{K0s}J. Takahashi {\it (STAR)}, SQM 2004, Sep. 15-20, 2004, Cape Town, South Africa.
\bibitem{xinthesis} Xin Dong, PH.D. Thesis, University of Science and
Technology of China, 2005.
\bibitem{johnson} J. Adams {\it et al.} (STAR Collaboration),
\Journal{\PRC}{70}{044902}{2004}; I.  Johnson, Ph.D. thesis,
U.C. Davis, 2002.
\bibitem{EMC}M. Beddo {\it et al.}, \Journal{\NIMA}{499}{725}{2003}.
\end{thebibliography}
\end{document}